\documentstyle[12pt]{article}

\newcommand{\be}{\begin{equation}}
\newcommand{\ee}{\end{equation}}
\newcommand{\bea}{\begin{eqnarray}}
\newcommand{\eea}{\end{eqnarray}}

\def\ep{\epsilon}
\def\lm{\lambda}

\def\ln{{\rm ln}}
\def\ep{\epsilon}

\begin{document}
\setlength{\baselineskip}{0.7cm}

\begin{titlepage}
\null
\begin{flushright}
hep-ph/0204069\\
UT-02-19\\
April, 2002
\end{flushright}
\vskip 1cm
\begin{center}
{\Large\bf 
(S)fermion Masses in Fat Brane Scenario
}

\lineskip .75em
\vskip 1.5cm

\normalsize

{\large Naoyuki Haba$^{a}$}
and {\large Nobuhito Maru$^{b}$}

\vspace{1cm}

{\it $^{a}$Faculty of Engneering, Mie University, Tsu, 
Mie, 514-8507, JAPAN} \\
{\it $^{b}$Department of Physics, University of Tokyo, 
Tokyo 113-0033, JAPAN} \\

\vspace*{10mm}

{\bf Abstract}\\[5mm]
{\parbox{13cm}{\hspace{5mm}
%
We discuss the fermion mass hierarchy and the flavor mixings 
in the fat brane scenario of a five dimensional SUSY theory. 
Assuming that the matter fields lives in the bulk, 
 their zero mode wave functions are Gaussians, 
 and Higgs fields are localized on the brane, 
 we find simple various types of the matter configurations generating 
 the mass matrices consistent with experimental data. 
Sfermion mass spectrum is also discussed 
 using the matter configurations found above. 
Which type of squark mass spectra (the degeneracy, the decoupling and 
 the alignment) is realized depends on the relative 
 locations of SUSY breaking brane 
 and the brane where Higgs fields are localized.}}

\end{center}

\end{titlepage}


\section{Introduction}
%
Utilizing extra dimensions has shed new insights into 
 various phenomenological aspects of the physics 
 in four dimensions. 
In particular, 
 Arkani-Hamed and Schmaltz have proposed an interesting mechanism, 
 which is refered to as ``fat brane scenario", 
 to generate small parameters naturally 
 in the effective four dimensional theories 
 from higher dimensional theories \cite{AS}. 
These small parameters are obtained 
 by a small overlap of wave functions, 
 even if the parameters in a fundamental theory are of order unity. 
This mechanism has been applied to various phenomenological issues so 
far, 
 such as the fermion mass hierarchy 
 \cite{MS,KT,Branco,Branco2,KT2,KY2}, 
 the doublet-triplet splitting \cite{KY,Maru,HM}, 
 and the sfermion mass generation \cite{KT,MSSS,KT2,HM2}.

In this paper, 
 we discuss the fermion mass hierarchy and the flavor mixings 
 in the fat brane scenario. 
As mentioned above, 
 some people has already considered this issue 
 \cite{MS,KT,Branco,Branco2,KT2,KY2}, 
 so it is necessary to clarify the difference 
 between our setup and the others. 
The setup of authors \cite{MS,Branco,Branco2} is as follows. 
A Higgs field lives in the bulk and its zero mode wave function 
 is flat. 
The matter fields also live in the bulk and its zero mode 
 wave functions are Gaussian and localized at different points 
 in extra dimensions. 
The fermion mass hierarchy is generated by 
 an overlap between the matter wave functions. 
The setup of authors \cite{KT,KY2} is that 
 the matter and Higgs live in the bulk and both zero mode 
 wave functions are Gaussian and are localized at different points 
 in extra dimensions. 
The fermion mass hierarchy is generated by 
 an overlap between the matter and Higgs wave functions. 
On the other hand, 
 our setup is slightly different from these two setups 
 in that Higgs fields are localized on a brane.\footnote{This setup 
 has also been considered in Ref.~\cite{KT2}. 
 The difference between their setup and ours is that 
 they considered that the form of the zero mode wave functions of 
 matter fields are exponential, but Gaussian in our case.} 
As for the matter fields, 
 our setup is the same as these two setups. 
The fermion mass hierarchy is determined by the values at a brane 
 where Higgs fields are localized. 
We find various types of the matter configurations 
 in a five dimensional supersymmetric (SUSY) theory, 
 which yield the fermion mass matrices consistent 
 with experimental data.

We also discuss the sfermion mass spectrum. 
Sfermion masses are generated by the overlap 
 between the wave functions of matter filefds and 
 the chiral superfileld with nonzero vacuum expectation value (VEV) 
 of the F-component localized on the SUSY breaking brane. 
If SUSY breaking is located far from the matter fields, 
 the sfermion mass spectrum is similar to the gaugino 
 mediation scenario since the gaugino masses receive 
 only the volume suppression but the sfermion masses receive 
 additional exponential suppression. 
In this case, the degeneracy solution to SUSY flavor problem is expected 
 as in the gaugino mediation. 
In Ref.~\cite{HM2}, 
 we have proposed that if SUSY breaking brane is located 
 between the first and the second generations, the sfermion mass 
 spectrum becomes the decoupling solution. 
But in \cite{HM2}, the flavor mixings were neglected in determining 
 the matter configuration for simplicity. 
In this paper, this point is improved. 
We estimate the sfermion masses and the mixings and see 
 whether the decoupling solution can be realized 
 when we take into account both of 
 the fermion masses and their flavor mixings. 
More interestingly, 
 if SUSY breaking is located 
 at the same point where Higgs fileds 
 are localized, the sfermion masses are 
 propotrtional to Yukawa couplings. 
In this case, 
 the alignment solution to SUSY flavor problem is expected. 
Thus, which type of the sfermion spectra is realized depends on 
 the relative location between a SUSY breaking brane and the brane 
 where Higgs fileds are localized in our framework.

This paper is organized as follows. 
In the next section, 
 we discuss the fermion mass hierarchy and the flavor mixings. 
We find various types of the matter configurations 
 consistent with the experimental data 
 in a five dimensional SUSY theory. 
In section 3, 
 corresponding to the fermion masses found above, 
 the sfermion mass spectrum is also discussed. 
Summary of our paper is given in the last section. 

\section{Fermion masses and mixings}
In this section, 
 we discuss the fermion masses and the flavor mixings 
 in a five dimensional SUSY theory. 
Let us consider the up-type Yukawa coupling, 
 for example,
%
\bea
W = \delta(y) \int dy Q_i(x,y) \bar{U}_j(x,y) H_u(x), 
\eea
where $x$ denotes the coordinate of 
 four dimensional Minkowski space-time, 
 $y$ is a fifth spatial coordinate of five dimensions. 
$i,j$ are the generation indices. 
The order one coefficient is implicit. 
$Q_i$ and $\bar{U}_i$ are the chiral superfield  
 which transform as $({\bf 3}, {\bf 2}, 1/6)$ 
 and $({\bf 3}^*, {\bf 1}, -2/3)$ 
 under the Standard Model (SM) gauge groups, 
 $SU(3)_C \times SU(2)_L \times U(1)_Y$. 
We assume here that the matter fields live in the bulk and 
 Higgs fields are localized on a brane at $y=0$. 
On a thick wall, 
 a zero mode wave function of the matter superfield 
 with the left (or right)-handed chirality is localized, 
 and the matter superfield with an opposite chirality 
 is not localized.\footnote{For 
 readers interested in the localization mechanism of the chiral 
 superfields, see an appendix in Ref.~\cite{KT}.} 
Integrating out the fifth dimensional degrees of freedom, 
 we obtain the effective Yukawa coupling in four dimensions 
 at the compactification scale as, 
%
\bea
\label{yukawa}
(y_{{\rm eff}})_{ij} \simeq {\rm exp}[-a^2 (y^2_{Q_i} 
+ y^2_{\bar{U}_j})], 
\eea
where we assume the form of the zero mode wave function of 
 the matter superfields to be Gaussian such as 
 ${\rm exp}[-a^2(y - y_{\Phi_i})^2]$, 
 where $M_*$ is the five dimensional Planck scale and 
 $a$ is the width of the matter zero mode wave functions. 
$y_{\Phi_i}$ is the coordinate where the matter superfield 
 $\Phi_i(=Q_i, \bar{U}_i, \bar{D}_i, ...)$ is localized. 
As is clear from (\ref{yukawa}), 
 the information of Yukawa hierarchy is interpreted as 
 the configuration of the matter fields in the 
 extra dimensions.

In the following subsections, 
 we consider various types of fermion mass matrices, 
 which well describe the fermion mass hierarchies and 
 the flavor mixings, 
 and discuss whether these mass matrices can be realized 
 from the fat brane point of view. 
Note that some of these mass matrices have been discussed 
 in the context of five or six dimensional SUSY orbifold 
 Grand Unified Theories (GUT) 
 \cite{HKSSU,HKS}. 

\subsection{``Anarchy type" fermion mass matrices}
At first, let us 
 consider the following matter configuration 
 in the fifth dimensional coordinate. 
%
\bea
\label{anarchy1}
\begin{array}{lll}
q_1^2 \simeq -2\ln \ep, & u_1^2 \simeq -2\ln \ep, 
& d_1^2 \simeq -\ln \ep, \\
q_2^2 \simeq -\ln \ep, & u_2^2 \simeq -\ln \ep, 
& d_2^2 \simeq -\ln \ep, \\
q_3^2 \simeq 0, & u_3^3 \simeq 0, & d_3^2 \simeq -\ln \ep, \\
l_1^2 \simeq 0, & e_1^2 \simeq -3\ln \ep, & n_1^2 \simeq 0, \\
l_2^2 \simeq 0, & e_2^2 \simeq -2\ln \ep, & n_2^2 \simeq 0, \\
l_3^2 \simeq 0, & e_3^2 \simeq -\ln \ep, & n_3^2 \simeq 0, \\
\end{array}
\eea
where $q_i$ means the coordinate in units of $a^{-1}$ 
 where $Q_i$ is localized, for instance. 
The configuration (\ref{anarchy1}) generates 
 the following mass matrices for up, down quark sectors 
 and the charged lepton sector:  
\bea
\label{anarchy3}
m_u \simeq 
\left(
\begin{array}{ccc}
\ep^4 & \ep^3 & \ep^2 \\
\ep^3 & \ep^2 & \ep \\
\ep^2 & \ep & 1
\end{array}
\right)\langle H_u \rangle,~
m_d \simeq 
\ep \left(
\begin{array}{ccc}
\ep^2 & \ep^2 & \ep^2 \\
\ep & \ep & \ep \\
1 & 1 & 1 \\
\end{array}
\right) \langle H_d \rangle, 
m_l \simeq 
\ep \left(
\begin{array}{ccc}
\ep^2 & \ep & 1 \\
\ep^2 & \ep & 1 \\
\ep^2 & \ep & 1
\end{array}
\right) \langle H_d \rangle,
\eea
where $\ep \simeq \lm^2$, and 
 $\lm$ is the Cabibbo angle, $\lm \simeq 0.2$. 
The above mass matrices suggest 
 the small tan$\beta$, and 
 generate the 
 fermion mass hierarchies, 
\bea
&&m_t:m_c:m_u \simeq 1:\ep^2:\ep^4, \\
&&m_b:m_s:m_d \simeq m_\tau:m_\mu:m_e \simeq 1: \ep :\ep^2. 
\eea
The mass matrices of the left-handed 
 neutrino and the right-handed neutrino sectors 
 are 
\bea
m_\nu^D \simeq 
\left(
\begin{array}{ccc}
1 & 1 & 1 \\
1 & 1 & 1 \\
1 & 1 & 1
\end{array}
\right) \langle H_d \rangle,~
m_N \simeq 
\left(
\begin{array}{ccc}
1 & 1 & 1 \\
1 & 1 & 1 \\
1 & 1 & 1
\end{array}
\right) M_R, 
\eea
where $M_R$ is around $10^{15-16}$ GeV. 
It is because the Majorana mass term in our model is assumed to be 
 generated by a nonzero VEV of the singlet $S$ field\footnote{
 The field $S$ have the lepton number in two units. 
 Thus its VEV might derive a massless Majoron, naively. 
 To discuss a Majoron is beyond the scope of this paper. 
 We simply asuume here, for example, the lepton 
 number is explicitly broken 
 in some sectors and avoid a massless Majoron.  } 
 localized on the brane at $y=0$ with the term, 
\bea
\delta(y) \frac{1}{2M_*} S (x) N_i (x, y) N_j(x, y). 
\eea
Integrating out the fifth dimensional degrees of freedom 
 and taking the VEV of $S$ to be $M_*$, 
 we obtain $M_R \simeq \sqrt{\frac{2}{\pi}}a^{-1} 
 \simeq 0.8 \times a^{-1}$. 
In the fat brane scenario, 
 $a^{-1}$ is at most $L_c^{-1} \simeq 10^{16}$GeV, 
 so $M_R \simeq 8.0 \times 10^{15}$ GeV. 
This is somewhat large compared to $M_R \simeq 10^{14}$GeV 
 which is the suitable magnitude for the 
 neutrino oscillation experiments, 
 but taking into account $O(1)$ coefficients 
 in neutrino mass matrices, 
 the value, $10^{15}$GeV, is sufficient. 
{}From these matrices, 
 the mass matrix of three light neutrinos $m_\nu^{(l)}$ 
 via see-saw mechanism \cite{seesaw} is obtained as 
\bea
m_\nu^{(l)} \simeq \frac{m_\nu^D (m_\nu^D)^t}{m_N} 
\simeq \left(
\begin{array}{ccc}
1 & 1 & 1 \\
1 & 1 & 1 \\
1 & 1 & 1 \\
\end{array}
\right)
\frac{\langle H_u \rangle^2}{M_R}. 
\eea
By diagonalizing these matrices, 
 the CKM \cite{CKM} and the MNS \cite{MNS} mixing matrices 
 are obtained, 
\bea
\label{CKM1}
V_{{\rm CKM}} \simeq 
\left(
\begin{array}{ccc}
1 & \ep & \ep^2 \\
\ep & 1 & \ep \\
\ep^2 & \ep & 1 \\
\end{array}
\right),~
V_{{\rm MNS}} \simeq 
\left(
\begin{array}{ccc}
1 & 1 & 1 \\
1 & 1 & 1 \\
1 & 1 & 1 \\
\end{array}
\right). 
\eea
They can naturally explain 
 why the flavor mixing in the quark sector 
 is small while the flavor mixing in the lepton sector is large 
 \cite{Babu,various,HaMu}. 
The above mass matrices and flavor mixings 
 are roughly consistent with the experimental data, 
 and explicit values of $O(1)$ 
 coefficients of mass matrices can really induce 
 the suitable magnitudes of fermion masses and 
 flavor mixing angles \cite{Babu}. 
The random coefficients have suggested 
 the probabilities of fermion masses and 
 flavor mixings \cite{HaMu}\cite{an2}. 
If we would like to obtain 
 the suitably large (small) Cabibbo angle 
 ($U_{e3}$), and small $m_{d,e}$,  
 of order the power of $\lambda$, 
 the matter configuration should be 
 modified. 
We will show two examples of the 
 improved fermion mass matrices 
 in the following two subsections. 

The configuration corresponding to the large 
 tan$\beta$ case is also 
  realized 
 by changing the locations of $\bar{D}_{1,2,3}$ 
 and $\bar{E}_{1,2,3}$ in (\ref{anarchy1}) as, 
\bea
\label{anarchy2}
\begin{array}{lll}
q_1^2 \simeq -2\ln \ep, & u_1^2 \simeq -2\ln \ep, & d_1^2 \simeq 0, \\
q_2^2 \simeq -\ln \ep, & u_2^2 \simeq -\ln \ep, & d_2^2 \simeq 0, \\
q_3^2 \simeq 0, & u_3^3 \simeq 0, & d_3^2 \simeq 0, \\
l_1^2 \simeq 0, & e_1^2 \simeq -2\ln \ep, & n_1^2 \simeq 0, \\
l_2^2 \simeq 0, & e_2^2 \simeq -\ln \ep, & n_2^2 \simeq 0, \\
l_3^2 \simeq 0, & e_3^2 \simeq 0, & n_3^2 \simeq 0. 
\end{array}
\eea
This configuration is more interesting than the previous configuration 
 since the localization patterns for $Q_i, \bar{U}_i, \bar{E}_i$ 
 and $\bar{D}_i, L_i$ are the same as for the generation 
 index. 
This implies that this case 
 can be embeded into the SU(5) GUT. 
The mass matrices for up, down quark sectors 
 and the charged lepton sector are given by 
%
\bea
\label{anarchy4}
m_u \simeq 
\left(
\begin{array}{ccc}
\ep^4 & \ep^3 & \ep^2 \\
\ep^3 & \ep^2 & \ep \\
\ep^2 & \ep & 1
\end{array}
\right)\langle H_u \rangle,~
m_d \simeq 
\left(
\begin{array}{ccc}
\ep^2 & \ep^2 & \ep^2 \\
\ep & \ep & \ep \\
1 & 1 & 1 \\
\end{array}
\right) \langle H_d \rangle, 
m_l \simeq 
\left(
\begin{array}{ccc}
\ep^2 & \ep & 1 \\
\ep^2 & \ep & 1 \\
\ep^2 & \ep & 1
\end{array}
\right) \langle H_d \rangle. 
\eea
Since the difference between the matrices (\ref{anarchy3}) 
 and (\ref{anarchy4}) only exists in 
 an overall factor, 
 the arguments for the fermion mass hierarchies and 
 the flavor mixings are the same 
 as 
 those of 
 the small tan$\beta$ case.

Here we would like to stress the importance of 
 the configurations found above. 
They are simpler than other configurations 
 in the literatures \cite{MS,KT,Branco,Branco2,KT2,KY2} 
 in the following points: 
\begin{enumerate}
 \item The configurations are obtained 
  in a {\em five dimensional theory}, \\
\vspace*{-7mm}
 \item {\em The width of the wave functions is common} to 
  all the matter fields, \\
\vspace*{-7mm}
 \item {\em The distribution of the matter superfields is very simple}. 
\end{enumerate}
%

\subsection{Improved mass matrices I}
{}From the view point of order estimation, 
 the mass matrices 
 discussed in the previous subsection 
 gives rather small Cabibbo angles and rather 
 large $U_{e3}$. 
As discussed in Ref.~\cite{HKS}, 
 there are two improvements of the ``Anarchy type" of 
 fermion mass matrices. 
In our framework, 
 these improvements 
 can be easily done by changing the locations 
 of $\bar{D}_1, L_1$ and $\bar{N}_1$ in (\ref{anarchy1}). 
Let us show the first example of the 
 improvements here. 
The configuration is given by 
\bea
\label{model11}
\begin{array}{lll}
q_1^2 \simeq -2\ln \ep, & u_1^2 \simeq -2\ln \ep, & d_1^2 
\simeq -2\ln \ep, \\
q_2^2 \simeq -\ln \ep, & u_2^2 \simeq -\ln \ep, & d_2^2 \simeq -\ln \ep, 
\\
q_3^2 \simeq 0, & u_3^3 \simeq 0, & d_3^2 \simeq - \ln \ep, \\
l_1^2 \simeq -\ln \ep, & e_1^2 \simeq -3\ln \ep, & n_1^2 
\simeq -\ln \ep, \\
l_2^2 \simeq 0, & e_2^2 \simeq -2\ln \ep, & n_2^2 \simeq 0, \\
l_3^2 \simeq 0, & e_3^2 \simeq -\ln \ep, & n_3^2 \simeq 0. \\
\end{array}
\eea
This configuration is simple enough 
 as that of ``Anarchy type". 
The mass matrices for up, down quark sectors and the charged 
 lepton sector from this configuration are 
\bea
\label{model13}
m_u \simeq 
\left(
\begin{array}{ccc}
\ep^4 & \ep^3 & \ep^2 \\
\ep^3 & \ep^2 & \ep \\
\ep^2 & \ep & 1
\end{array}
\right) \langle H_u \rangle,~
m_d \simeq 
\ep \left(
\begin{array}{ccc}
\ep^3 & \ep^2 & \ep^2 \\
\ep^2 & \ep & \ep \\
\ep & 1 & 1
\end{array}
\right) \langle H_d \rangle,~
m_l \simeq 
\ep \left(
\begin{array}{ccc}
\ep^3 & \ep^2 & \ep \\
\ep^2 & \ep & 1 \\
\ep^2 & \ep & 1
\end{array}
\right) \langle H_d \rangle,
\eea
The mass hierarchies in the down quark sector 
 and the charged lepton sector are modified as 
\bea
\label{hier1}
&&m_t:m_c:m_u \simeq 1:\ep^2:\ep^4, \\
&&m_b:m_s:m_d \simeq m_\tau:m_\mu:m_e \simeq 1: \ep :\ep^3,
\eea
This suggests the suitable order of magnitudes of $m_e$ and 
 $m_d$.

The mass matrices of 
 the left-handed neutrino sector and the right-handed neutrino sector 
 are given by 
\bea
\label{model12}
m_\nu^D \simeq 
\left(
\begin{array}{ccc}
\ep^2 & \ep & \ep \\
\ep & 1 & 1 \\
\ep & 1 & 1
\end{array}
\right) \langle H_d \rangle,~
m_N \simeq 
\left(
\begin{array}{ccc}
\ep^2 & \ep & \ep \\
\ep & 1 & 1 \\
\ep & 1 & 1
\end{array}
\right) M_R. 
\eea
The mass matrix of three light neutrinos $m^{(l)}_\nu$ 
 via see-saw mechanism is given by 
\bea
m_\nu^{(l)} \simeq \frac{m_\nu^D (m_\nu^D)^t}{m_N} 
\simeq \left(
\begin{array}{ccc}
\ep^2 & \ep & \ep \\
\ep & 1 & 1 \\
\ep & 1 & 1 \\
\end{array}
\right)
\frac{\langle H_u \rangle^2}{M_R}. 
\eea
%
Here we assume the rank 
 of $2 \times 2$ submatrix for the second and the third 
 generations in $m^{(l)}_\nu$ is reduced. 
Otherwise, 
 the large mixing between the first and the second generations 
 cannot be realized. 
When this rank reduction of the submatrix is assumed 
 in (17), 
 the CKM and the MNS matrices are obtained as 
\bea
V_{{\rm CKM}} \simeq 
\left(
\begin{array}{ccc}
1 & \ep & \ep^2 \\
\ep & 1 & \ep \\
\ep^2 & \ep & 1 \\
\end{array}
\right),~
V_{{\rm MNS}} \simeq 
\left(
\begin{array}{ccc}
1/\sqrt{2} & 1/\sqrt{2} & \ep \\
1/2 & -1/2 & 1/\sqrt{2} \\
-1/2 & 1/2 & 1/\sqrt{2} \\
\end{array}
\right). 
\eea
This type of the MNS matrix is so-called bi-maximal one.

The configuration corresponding 
 to the large tan$\beta$ case is easily obtained 
 by changing the locations of 
 $\bar{D}_{1,2,3}$ and $\bar{E}_{1,2,3}$ in (\ref{model11}) as 
\bea
\label{model12}
\begin{array}{lll}
q_1^2 \simeq -2\ln \ep, & u_1^2 \simeq -2\ln \ep, & d_1^2 
\simeq -\ln \ep, \\
q_2^2 \simeq -\ln \ep, & u_2^2 \simeq -\ln \ep, & d_2^2 \simeq 0, \\
q_3^2 \simeq 0, & u_3^3 \simeq 0, & d_3^2 \simeq 0, \\
l_1^2 \simeq -\ln \ep, & e_1^2 \simeq -2\ln \ep, & n_1^2 
\simeq -\ln \ep, \\
l_2^2 \simeq 0, & e_2^2 \simeq -\ln \ep, & n_2^2 \simeq 0, \\
l_3^2 \simeq 0, & e_3^2 \simeq 0, & n_3^2 \simeq 0. \\
\end{array}
\eea
The corresponding mass matrices of the up and down quark sectors 
 and the charged lepton sector are 
\bea
\label{model14}
m_u \simeq 
\left(
\begin{array}{ccc}
\ep^4 & \ep^3 & \ep^2 \\
\ep^3 & \ep^2 & \ep \\
\ep^2 & \ep & 1
\end{array}
\right) \langle H_u \rangle,~
m_d \simeq 
\left(
\begin{array}{ccc}
\ep^3 & \ep^2 & \ep^2 \\
\ep^2 & \ep & \ep \\
\ep & 1 & 1
\end{array}
\right) \langle H_d \rangle,~
m_l \simeq 
\left(
\begin{array}{ccc}
\ep^3 & \ep^2 & \ep \\
\ep^2 & \ep & 1 \\
\ep^2 & \ep & 1
\end{array}
\right) \langle H_d \rangle. 
\eea
This case can be embeded into the SU(5) GUT. 
The results for the fermion mass hierarchies 
 and the flavor mixings are 
 the same as those of the small tan$\beta$ case.

\subsection{Improved mass matrices II}
The second example of the improvements 
 shows the changes of 
 the locations of $\bar{D}_{1,2,3}, L_{1,2,3}$ 
 and $\bar{N}_{1,2,3}$ in (\ref{anarchy1}). 
The matter configuration is 
%
\bea
\label{model21}
\begin{array}{lll}
q_1^2 \simeq -2\ln \ep, & u_1^2 \simeq -2\ln \ep, 
& d_1^2 \simeq -3 \ln \ep, \\
q_2^2 \simeq -\ln \ep, & u_2^2 \simeq -\ln \ep, 
& d_2^2 \simeq -2\ln \ep, \\
q_3^2 \simeq 0, & u_3^3 \simeq 0, & d_3^2 \simeq -2\ln \ep, \\
l_1^2 \simeq -2\ln \ep, & e_1^2 \simeq -3\ln \ep, 
& n_1^2 \simeq -2\ln \ep, \\
l_2^2 \simeq -\ln \ep, & e_2^2 \simeq -2\ln \ep, 
& n_2^2 \simeq -\ln \ep, \\
l_3^2 \simeq -\ln \ep, & e_3^2 \simeq -\ln \ep, 
& n_3^2 \simeq -\ln \ep. \\
\end{array}
\eea
%
The resulting mass matrices for up, down quark sectors and the charged 
 lepton sector are 
\bea
\label{model21}
m_u \simeq 
\left(
\begin{array}{ccc}
\ep^4 & \ep^3 & \ep^2 \\
\ep^3 & \ep^2 & \ep \\
\ep^2 & \ep & 1
\end{array}
\right) \langle H_u \rangle,~
m_d \simeq 
\ep^2 \left(
\begin{array}{ccc}
\ep^3 & \ep^2 & \ep^2 \\
\ep^2 & \ep & \ep \\
\ep & 1 & 1
\end{array}
\right) \langle H_d \rangle,~
m_l \simeq 
\ep^2 \left(
\begin{array}{ccc}
\ep^3 & \ep^2 & \ep \\
\ep^2 & \ep & 1 \\
\ep^2 & \ep & 1
\end{array}
\right) \langle H_d \rangle,
\eea
and those of the left-handed neutrino sector and the right-handed 
 neutrino sector are given by 
\bea
m_\nu^D \simeq \ep^2~
\left(
\begin{array}{ccc}
\ep^2 & \ep & \ep \\
\ep & 1 & 1 \\
\ep & 1 & 1 \\
\end{array}
\right) \langle H_d \rangle,~
m_N \simeq \ep^2~
\left(
\begin{array}{ccc}
\ep^2 & \ep & \ep \\
\ep & 1 & 1 \\
\ep & 1 & 1 \\
\end{array}
\right) M_R. 
\eea
Then, the mass matrix of three light neutrinos $m_\nu^{(l)}$ 
 via the see-saw mechanism is given by 
\bea
m_\nu^{(l)} \simeq \frac{m_\nu^D (m_\nu^D)^t}{m_N} 
\simeq \ep^2~\left(
\begin{array}{ccc}
\ep^2 & \ep & \ep \\
\ep & 1 & 1 \\
\ep & 1 & 1 \\
\end{array}
\right)
\frac{\langle H_u \rangle^2}{M_R}. 
\eea
Above mass matrices are the same as those of improvement I 
 except for 
 the overall factors. 
The fermion mass hierarchies of quarks and leptons are 
 the same as those of improvement I. 
The flavor mixing matrices, $V_{{\rm CKM}}$ and $V_{{\rm MNS}}$, 
 are also the same as those of the improvement I. 
The difference between improvement I and II 
 exists in the value of 
 tan$\beta$. 
Compairing to the improvement I, 
 the 
 small tan$\beta$ is prefer in 
 the improvement II. 

We can also find another configuration 
 by changing the locations of 
 $\bar{D}_{1,2,3}$ and $\bar{E}_{1,2,3}$ in (\ref{model21}), 
\bea
\label{model22}
\begin{array}{lll}
q_1^2 \simeq -2\ln \ep, & u_1^2 \simeq -2\ln \ep, 
& d_1^2 \simeq -2\ln \ep, \\
q_2^2 \simeq -\ln \ep, & u_2^2 \simeq -\ln \ep, 
& d_2^2 \simeq -\ln \ep, \\
q_3^2 \simeq 0, & u_3^3 \simeq 0, & d_3^2 \simeq -\ln \ep, \\
l_1^2 \simeq -\ln \ep, & e_1^2 \simeq -3\ln \ep, 
& n_1^2 \simeq -\ln \ep, \\
l_2^2 \simeq 0, & e_2^2 \simeq -2\ln \ep, & n_2^2 \simeq 0, \\
l_3^2 \simeq 0, & e_3^2 \simeq - \ln \ep, & n_3^2 \simeq 0. \\
\end{array}
\eea
This configuration is nothing but that of the improvement I
(\ref{model11}).

\section{Sfermion Masses}
In this section, 
 we would like to discuss the sfermion mass spectrum, 
 especially the squark mass spectrum. 
In extra dimensions, 
 the sfermion masses correlate with the fermion masses 
 by introducing SUSY breaking brane 
 because the sfermion masses are determined by the overlap of 
 wave functions of the matter fields and the chiral superfields 
 with nonvanishing F-term VEV on SUSY breaking brane. 
Since we have determined the matter configuration consistent 
 with experimental data in the previous section, 
 the sfermion mass spectrum can be calculated and predicted. 

The gaugino masses are generated at tree level 
 since we assume that the gauge supermultiplets live 
 in a thick wall throughout this paper, 
%
\bea
\label{gmass}
\delta(y-L) \int d^2 \theta \frac{X(x)}{M_*^2} 
W^{\alpha}(x,y) W_{\alpha}(x,y) 
\Rightarrow M_{\lambda} = \frac{F}{M_*^2 L_c}, 
\eea
where $W_\alpha$ is the field strength tensor superfield and 
 $L_c$ is the width of the thick wall which should be considered as 
 the compactification length in our framework. 
The gaugino masses should be around 100 GeV, 
 so we obtain 
\bea
\label{gaugino}
\frac{F}{M_*} \simeq 100(M_* L_c). 
\eea
%

%
%
%
%

In our previous paper, 
 we have proposed the mechanism to generate the sfermion masses 
 in the fat brane scenario \cite{HM2}. 
Let us briefly review the scenario. 
A brane, we call ``SUSY breaking brane", is introduced at $y=L$, 
 where the chiral superfields $X$ with nonvanishing F-term VEV 
 ($X=\theta^2 F$) is assumed to be localized. 
The extra vector-like superfields $\Phi'$, $\bar{\Phi}'$ 
 with mass $M < M_*$ are also introduced and 
 assumed to be localized on a SUSY breaking brane. 
We consider here the following superpotential, 
\bea
W &=& \delta(y-L) \int dy [\frac{\lm}{\sqrt{M_*}} X(x) 
\Phi_i(x,y) \bar{\Phi}'(x) 
+ M \Phi(x)' \bar{\Phi}'(x)], \\
&=& \frac{\lm}{\sqrt{M_* a}} 
{\rm exp}[-a^2(L-y_{\Phi_i})^2] X(x) \Phi_i(x) 
\bar{\Phi}'(x) + M \Phi'(x) \bar{\Phi}'(x), 
\eea
where $\lm$ is a dimensionless constant of order unity. 
Below the scale $M$, 
 we can integrate out the massive superfields $\Phi'$ and 
 $\bar{\Phi}'$, then the superpotential vanishes and 
 the effective K\"ahler potential receives the correction 
 at tree level, 
\bea
\label{effK}
\delta K = \frac{1}{\sqrt{M_* a}} \frac{1}{M^2} {\rm exp}
[-a^2 \{ (L-y_{\Phi_i})^2 + (L-y_{\Phi_j})^2 \}] 
X^\dag X \Phi_i^\dag \Phi_j. 
\eea
The sfermion masses coming from (\ref{effK}) 
 at the compactification scale are 
\bea
\label{sfermion}
\tilde{m}^2_{ij} \simeq \frac{1}{\sqrt{M_* a}} {\rm exp}
[-a^2 \{ (L-y_{\Phi_i})^2 + (L-y_{\Phi_j})^2 \}] 
\frac{|F|^2}{M^2}.  
\eea
It is crucial that the scale suppressing the K\"ahler potential 
 is replaced with $M < M_*$ not so as to be negligibly 
 small.\footnote{Without 
 introducing the extra vector-like superfields, the sfermion masses are 
 negligibly small due to the exponential suppression \cite{HM2}.} 
Note that $F<M^2$ is assumed in this argument and 
 also the overall sign of the K\"ahler potential is assumed 
 to be positive.

Before discussing the sfermion masses, 
 we would like to mention A-terms in our scenario. 
In our scenario, A-terms are induced, for instance, 
 from the term 
\bea
\int d^2 \theta \frac{X}{M_*^2} Q_i \bar{U}_j H_u. 
\eea
By integrating out the fifth dimensional degrees of freedom, 
 it turns out immediately that A-terms are vanishing 
 unless the SUSY breaking brane is located at the same point 
 where Higgs fields are localized. 
Nonvanishing A-parameters are expressed by 
\bea
A = \frac{F}{M_*^2 a} (y_{{\rm eff}})_{ij}. 
\eea
It is remarkable that in our framework A-parameters are 
 necessarily proportional to Yukawa coupling constants.

If SUSY breaking brane is located apart from the brane where 
 Higgs fields are localized, the sfermion masses 
 (\ref{sfermion}) are negligibly small compared to the gaugino masses 
 (\ref{gmass}) since the sfermion masses receive the additional 
 exponential suppression factor. 
This boundary condition is similar to the gaugino mediation scenario 
 \cite{gMSB}. 
Therefore, the degeneracy solution to SUSY flavor problem is expected 
 as in the gaugino mediation scenario.

In \cite{HM2}, we have proposed 
 the decoupling solution in the fat brane scenatio. 
Now, we shall discuss this scenario. 
If a pair of vector-like superfields are introduced 
 for each matter chiral superfield $Q_i, \bar{U}_i, \bar{D}_i...$, 
 we obtain the K\"ahler potential with no flavor mixings. 
We consider here this case, for simplicity. 
Sfermion masses generating from the K\"ahler potential 
 with no flavor mixings are 
%
\bea
\tilde{m}_i^2 = ({\rm exp}[-a^2(L-y_{\Phi_i})^2])^2 
\frac{1}{M_* a} \frac{|F|^2}{M^2}. 
\eea
%

Now, we apply this mechanism to three types of the matter 
 configuration found 
 in the previous section and discuss whether 
 the decoupling solution can be realized or not. 
Let us begin with the ``Anarchy" model. 
Since the decoupling solution requires that the masses of the first 
 and the second generations should be of 
 ${\cal O}(10{\rm TeV})$ and that of the third generation be around 
 ${\cal O}(100{\rm GeV})$ for naturalness, namely, 
\bea
\label{3gen}
&&\tilde{m}_{Q_3} \simeq \tilde{m}_{\bar{U}_3} 
\simeq 100~{\rm GeV} 
\Rightarrow \frac{1}{\sqrt{M_* a}} 
\frac{F}{M}~{\rm exp}[-(a L)^2] \simeq 100~{\rm GeV}, \\
\label{2gen}
&&\tilde{m}_{Q_1} \simeq \tilde{m}_{\bar{U}_1} \simeq 10~{\rm TeV} 
\Rightarrow \frac{1}{\sqrt{M_* a}} 
\frac{F}{M}~{\rm exp}[-(a L - \sqrt{-2{\rm ln}\ep})^2] 
\simeq 10~{\rm TeV}. 
\eea
Using Eqs.~(\ref{gaugino}) and (\ref{3gen}), 
\bea
\label{mass}
M \simeq \frac{(M_* L_c) M_*}{\sqrt{M_* a}} 
{\rm exp} [-(a L)^2]
\eea
is obtained, 
{}From Eqs.~(\ref{3gen}) and (\ref{2gen}), 
\bea
\label{l}
a L \simeq 2.22 
\eea
is obtained. 
Eq.~(\ref{mass}) can be further rewritten by using Eq.~(\ref{l}) as 
\bea
M \simeq 7.18 \times 10^{-3}~\frac{(M_* L_c) M_*}{\sqrt{M_* a}}. 
\eea
Taking $M_* \simeq 10^{18}$ GeV, 
 $L_c^{-1} \simeq a^{-1} \simeq 10^{16}$ GeV, 
 various scales in our theory are obtained, 
\bea
L^{-1} \simeq 4.50 \times 10^{17}~{\rm GeV},~
M \simeq 7.18 \times 10^{16}~{\rm GeV},~
\sqrt{F} \simeq 10^{11}~{\rm GeV}. 
\eea
As mentioned above, 
 $F < M^2$ is satisfied. 
Other sfermion masses are calculated, 
\bea
\tilde{m}_{Q_2} \simeq \tilde{m}_{\bar{U}_2} \simeq 
\tilde{m}_{D_{1,2,3}} 
\simeq 10^{-1} \times \frac{10^{22}}{7.18 \times 10^{16}} 
{\rm exp}[-(2.221756-1.977883)^2] 
\simeq 13.12~{\rm TeV}. 
\eea

We note that we have an another choice 
for the decoupling solution 
 requirement, namely, $\tilde{m}_{Q_2} \simeq 
 \tilde{m}_{\bar{U}_2} \simeq \tilde{m}_{D_{1,2,3}} 
 \simeq 10~{\rm TeV}$. 
In this case, 
 the squark masses of the first generation $Q_1$ and $\bar{U}_1$ 
 become 3.8 TeV, which is smaller than 10 TeV. 
Therefore, this choice is not plausible for the decoupling solution.

As for the sfermion mass spectrum of Improvement I,  
 they are almost same as that of ``Anarchy" model 
 except for the right-handed sdown mass 
 because the difference in the quark sector 
 between two configurations 
 Eqs.~(\ref{anarchy1}) and (\ref{model11})  is 
 the location of the right-handed down chiral superfield. 
As for the Improvement II, 
 the differences in the quark sector 
 between the ``Anarchy type" and the Improvement II   
 are 
 the locations of $\bar{D}_{1,2,3}$. 
In particular, 
 since $\bar{D}_1$ is localized a little apart from a SUSY breaking 
 brane, the mass of $\bar{D}_1$ becomes slightly small. 
The sfermion masses spectra are summarized in the Table 1. 
\begin{table}
\begin{center}
 \begin{tabular}{|l|l|l|}
 \hline
 {\rm ``Anarchy" type}& {\rm Improvement~1} & {\rm Improvement~2} \\
 \hline
 $\tilde{m}_{Q_3,\bar{U}_3} \simeq 100~{\rm GeV}$ & 
 $\tilde{m}_{Q_3,\bar{U}_3} \simeq 100~{\rm GeV}$ & 
 $\tilde{m}_{Q_3,\bar{U}_3} \simeq 100~{\rm GeV}$ \\
 & & $\tilde{m}_{\bar{D}_1} \simeq 3.27~{\rm TeV}$ \\ 
 $\tilde{m}_{Q_2,\bar{U}_2, \bar{D}_{1,2,3}} \simeq 13.12~{\rm TeV}$ & 
 $\tilde{m}_{Q_2,\bar{U}_2,\bar{D}_{2,3}} \simeq 13.12~{\rm TeV}$ & 
 $\tilde{m}_{Q_2,\bar{U}_2} \simeq 13.12~{\rm TeV}$ \\
 $\tilde{m}_{Q_1,\bar{U}_1} \simeq 10~{\rm TeV}$ & 
 $\tilde{m}_{Q_1,\bar{U}_1,\bar{D}_1} \simeq 10~{\rm TeV}$ & 
 $\tilde{m}_{Q_1,\bar{U}_1,\bar{D}_{2,3}} \simeq 10~{\rm TeV}$\\
 \hline
 \end{tabular}
\end{center}
\caption{Sfermion mass spectra of three models}
\end{table}

We give some comments on the sfermion mass spectrum. 
First, as seen in Ref.~\cite{HM2}, 
 the mass of the right-handed sbottom $\tilde{m}_{\bar{D}_3}$ is 
 around 10 TeV for all models, 
 which is somewhat large from the viewpoint of 
 the decoupling solution. 
This means that the large tan$\beta$ is preferable. 
Second, the mass of the right-handed sdown 
 $\tilde{m}_{\bar{D}_1}$ is somewhat small in Improvement II 
 because the right-handed down chiral superfield $\bar{D}_1$ is 
 localized far apart from the SUSY breaking brane compared to 
 the location where other matter chiral superfields of the first 
 and the second generations. 
Third, 
 since these sfermion mass spectrum is determined at 
 the compactification scale, the third generation sfermion mass 
 squareds might be driven negative at the weak scale 
 through the two-loop renormalization group effects of 
 the heavy first-two generation sfermion masses \cite{negative}. 
To avoid this, 
 we have to add the extra fields with negative SUSY breaking 
 mass squareds \cite{HKN}. 
We do not discuss this point in detail in this paper. 
Finally, 
 the sfermion mass spectrum from the fermion masses for 
 (\ref{anarchy2}) and (\ref{model12}) are not viable as 
 the decoupling solutions because $\tilde{m}_{\bar{D}_{1,2}}$ or 
 $\tilde{m}_{\bar{D}_2}$ becomes the order of 100 GeV.



Another interesting sfermion spectrum can be obtained 
 if the SUSY breaking brane is put on the same point 
 where the Higgs fields are localized and 
 the vector-like superfields are introduced 
 on the SUSY breaking brane. 
%
%
In ``Anarchy" model with small tan$\beta$, for instance,  
 we obtain the squark mass matrices 
%
\bea
\label{mumd}
&&M^2_u = \left( 
\begin{array}{cc}
\tilde{m}^2_Q + m_u^2 & m_u (a_u - \mu {\rm cot}\beta) \\
m_u (a_u - \mu {\rm cot}\beta) & \tilde{m}^2_{\bar{U}} + m_u^2
\end{array}
\right), \\
&&M^2_d = \left( 
\begin{array}{cc}
\tilde{m}^2_Q + m_d^2 & m_d (a_d - \mu {\rm tan}\beta) \\
m_d (a_d - \mu {\rm tan}\beta) & \tilde{m}^2_{\bar{D}} + m_d^2
\end{array}
\right), 
\eea
where 
\bea
&&\tilde{m}^2_{Q} \simeq \tilde{m}^2_{\bar{U}} 
\simeq 
\left(
\begin{array}{ccc}
\ep^4 & \ep^3 & \ep^2 \\
\ep^3 & \ep^2 & \ep \\
\ep^2 & \ep & 1
\end{array}
\right) \frac{1}{M_* a} 
\left( \frac{F}{M} \right)^2,~
\tilde{m}^2_{\bar{D}} \simeq 
\ep^2 \left(
\begin{array}{ccc}
1 & 1 & 1 \\
1 & 1 & 1 \\
1 & 1 & 1 \\
\end{array}
\right) \frac{1}{M_* a} 
\left( \frac{F}{M} \right)^2, \\ 
&&a_{u,d} \equiv A_{u,d}/y_{u,d} \simeq 
\frac{1}{M_* a} \frac{F}{M}, 
\eea
%
%
%
and $m_u$, $m_d$ are the fermion mass matrices 
 in Eqs.~(\ref{anarchy3}). 
Also, 
 we omitted the elements coming from $SU(2)_L \times U(1)_Y$ 
 D-terms since the discussion below is not changed. 
We shall discuss the possibility of the realization of 
 the alignment solution\footnote{In order to be 
 preserve the alignment solution against the quantum corrections, 
 an additional mechanism might be needed in our scenario. }
 to SUSY flavor problem \cite{NS}. 
In the sfermion martrices $M^2_u, M^2_d$, 
 if the off-diagonal elements are smaller than the diagonal ones, 
 we can obtain the alignment solution 
 since both of the sfermion mass squareds and 
 the fermion mass squareds in each diagonal element 
 take the same form and these can be simultaneously diagonalized. 
In the above case, 
 the dominant terms in the diagonal elements are 
 $\tilde{m}^2_Q, \tilde{m}^2_{\bar{U}}$ 
 and $\tilde{m}^2_{\bar{D}}$. 
As for the off-diagonal elements, 
 we note that 
 ${\rm tan}\beta \simeq \ep m_t/m_b \simeq 10^{-1}$ 
 in the present case. 
Assuming that the sfermion masses 
 $\tilde{m}^2_{Q,\bar{U},\bar{D}}$ 
 are around 1 TeV, 
 we need the constraint from $\tilde{m}^2_{Q, \bar{U}}$ 
\bea
\frac{\ep^2}{\sqrt{M_* a}} \left( \frac{F}{M} \right) 
\simeq 1 {\rm TeV}. 
\eea
This can be rewritten as 
\bea
\label{a}
a_{u,d} \simeq \frac{10^7}{\sqrt{M_* a}}~{\rm GeV}. 
\eea
Assuming that $\mu \simeq 100~{\rm GeV}$, 
\bea
\label{b}
\mu~{\rm tan}\beta \simeq 10~{\rm GeV},
~\mu~{\rm cot}\beta \simeq 1~{\rm TeV} 
\eea
are obtained. 
Which term dominates in the off-diagonal elements depends on 
 the value of $M_* a$ in (\ref{a}). 
If 
 $a_{u,d}$ are dominant, 
 the conditions to obtain the alignment solution are 
\bea
\tilde{m}^2_{Q,\bar{U}} > m_u a_u,~\tilde{m}^2_{Q,\bar{D}} > m_d a_d. 
\eea
Most stringent constraint coming from the top quark sector is 
\bea
1~{\rm TeV^2} > \frac{10^9}{\sqrt{M_* a}}~{\rm GeV^2} 
\Rightarrow M_*a >10^6, 
\eea
which provides a lower bound of the width of the wave functions.


The slepton mass matrix is 
\bea
M^2_e = \left( 
\begin{array}{cc}
\tilde{m}^2_L + m_l^2 & m_l (a_e - \mu {\rm tan}\beta) \\
m_l (a_e - \mu {\rm tan}\beta) & \tilde{m}^2_{\bar{E}} + m_l^2
\end{array}
\right), 
\eea
where 
\bea
&&\tilde{m}_L^2 
\simeq \left(
\begin{array}{ccc}
1 & 1 & 1 \\
1 & 1 & 1 \\
1 & 1 & 1 \\
\end{array}
\right) \frac{1}{M_* a} 
\left( \frac{F}{M} \right)^2,~
\tilde{m}^2_{\bar{E}} \simeq 
\ep^2 \left(
\begin{array}{ccc}
\ep^4 & \ep^3 & \ep^2 \\
\ep^3 & \ep^2 & \ep \\
\ep^2 & \ep & 1
\end{array}
\right) \frac{1}{M_* a} 
\left( \frac{F}{M} \right)^2, \\
&&a_e \equiv A_e/y_e \simeq 
\frac{1}{M_* a} \frac{F}{M}, 
\eea
and $m_l$ is given in (\ref{anarchy3}). 

The squark mass matrices with the large tan$\beta$ case 
 are written by 
\bea
&&\tilde{m}^2_{Q} \simeq \tilde{m}^2_{\bar{U}} \simeq 
\left(
\begin{array}{ccc}
\ep^4 & \ep^3 & \ep^2 \\
\ep^3 & \ep^2 & \ep \\
\ep^2 & \ep & 1
\end{array}
\right) \frac{1}{M_* a} 
\left( \frac{F}{M} \right)^2,~
\tilde{m}^2_{\bar{D}} \simeq 
\left(
\begin{array}{ccc}
1 & 1 & 1 \\
1 & 1 & 1 \\
1 & 1 & 1 \\
\end{array}
\right) \frac{1}{M_* a} 
\left( \frac{F}{M} \right)^2, 
\eea
$a_{u,d}$ are the same as in the small tan$\beta$ case 
 and $m_u$ and $m_d$ are the fermion mass matrices 
 in Eqs.~(\ref{anarchy4}). 
We can repeat the discussion on the alignment solution. 
In the large tan$\beta$ case. 
 the constraint for the width of the wave functions becomes 
 $M_* a > 10^{-2}$. 
This constraint is always satisfied 
 since $a > M_*^{-1}$ in the fat brane scenario. 

 

%
%

The slepton mass matrix is 
\bea
&&\tilde{m}_L^2 
\simeq \left(
\begin{array}{ccc}
1 & 1 & 1 \\
1 & 1 & 1 \\
1 & 1 & 1 \\
\end{array}
\right) \frac{1}{M_* a} 
\left( \frac{F}{M} \right)^2,~
\tilde{m}^2_{\bar{E}} \simeq 
\left(
\begin{array}{ccc}
\ep^4 & \ep^3 & \ep^2 \\
\ep^3 & \ep^2 & \ep \\
\ep^2 & \ep & 1
\end{array}
\right) \frac{1}{M_* a} 
\left( \frac{F}{M} \right)^2, 
\eea
$a_e$ is the same as the small tan$\beta$ case 
and $m_l$ is given in (\ref{anarchy4}).

It is important to note that which type of 
 the sfermion mass spectrum is realized depends on 
 the relative location between the SUSY breaking brane 
 and the brane where Higgs fields are localized. 
Thus, our framework provide a unified picture of the mechanism 
 for generating the sfermion masses.

\section{Summary}
We have discussed the fermion masses and the mixings 
 in the fat brane scenario of a five dimensional SUSY theory. 
In extra dimensions, 
 the information of Yukawa hierarchy is interpreted as 
 that of the location where the matter fields are localized. 
There have been many configurations consistent with the experimental 
 data, which corresponds to the way to embed the matter 
 fields and Higgs fields in extra dimensions. 
We have considered the case in this paper 
 that the matter fields live in the bulk 
 and Higgs fields are localized on the brane. 
We have found various types of the matter configurations 
 which yields the mass matrices consistent with 
 the fermion mass hierarchy and the flavor mixings. 
The configurations found in this paper are very simple 
 compared to others in the literature 
 \cite{MS,KT,Branco,Branco2,KT2,KY2} in that 
\begin{enumerate}
\item The configurations are obtained 
 in a five dimensional theory, \\
\vspace*{-7mm}
 \item The width of the wave functions is common to 
  all the matter fields, \\
\vspace*{-7mm}
 \item The distribution of the matter superfields is very simple. 
\end{enumerate}
%


We have also discussed three types of the sfermion spectrum 
 in our framework. 
If SUSY breaking brane is located far apart from the region where 
 the matter fields are localized, the sfermion mass spectrum is 
 similar to the gaugino mediation scenario \cite{gMSB}. 
The degeneracy solution to SUSY flavor problem is expected. 
If SUSY breaking brane is located between the points where 
 the first and the second generations are localized, 
 the decoupling spectrum proposed in \cite{HM2} is realized. 
If the SUSY breaking brane is located 
 at the same point where Higgs fields are localized, 
 the sfermion masses are proportional to Yukawa couplings. 
This case might suggests the alignment solution \cite{NS} 
 to SUSY flavor problem. 
We emphasize that our framework provides a unified 
 picture of the mechanism for generating the sfermion masses. 
Which scenario is realized depends on the relative location 
 of the SUSY breaking brane and 
 the brane where Higgs fields are localized.

As a future direction, 
 it is interesting to study in detail 
 the slepton mass spectrum and their mixings in our framework. 

\vspace*{10mm}


\begin{flushleft}
{\Large\bf Acknowledgments}
\end{flushleft}
N.H. is supported by the Grant-in-Aid for Scientific Research, 
Ministry of Education, Science and Culture, Japan (No.12740146) and 
N.M. is supported 
by the Japan Society for the Promotion of Science 
for Young Scientists (No.08557).

\vspace{1cm}
     %

\end{document}